\documentclass[aps,pra,amsmath,showpacs,twocolumn,amssymb,superscriptaddress]{revtex4}

\usepackage[dvips]{graphicx}
\usepackage{dcolumn}
\usepackage{bm}
\usepackage{dsfont}
\usepackage{color}

\usepackage{epstopdf}

\usepackage{url}
\usepackage[colorlinks=true ,citecolor=blue,urlcolor=blue]{hyperref}
\usepackage{amsmath,amssymb}

\def\bs#1{\boldsymbol{#1}}
\def\txt#1{\textnormal{#1}}

\begin{document}


\title{Detecting Chiral Edge States in the Hofstadter Optical Lattice}

\author{Nathan Goldman}
\email{ngoldmanATulb.ac.be}
\affiliation{Center for Nonlinear Phenomena and Complex Systems - Universit\'e Libre de Bruxelles (U.L.B.), B-1050 Brussels, Belgium}
\author{J\'er\^ome Beugnon}
\affiliation{Laboratoire Kastler Brossel, CNRS, ENS, UPMC, 24 rue Lhomond, 75005 Paris}
\author{Fabrice Gerbier}
\affiliation{Laboratoire Kastler Brossel, CNRS, ENS, UPMC, 24 rue Lhomond, 75005 Paris}


\begin{abstract} 
We propose a realistic scheme to detect topological edge states in an optical lattice subjected to a synthetic magnetic field, based on a generalization of Bragg spectroscopy sensitive to angular momentum. We demonstrate that using a well-designed laser probe, the Bragg spectra provide an unambiguous signature of the topological edge states that establishes their chiral nature. This signature is present for a variety of boundaries, from a hard wall to a smooth harmonic potential added on top of the optical lattice. Experimentally, the Bragg signal should be very weak. To make it detectable, we introduce a ``shelving method", based on Raman transitions, which transfers angular momentum and changes the internal atomic state simultaneously. This scheme allows to detect the weak signal from the selected edge states on a dark background, and drastically improves the detectivity. It also leads to the possibility to directly \emph{visualize} the topological edge states, using \emph{in situ} imaging, offering a unique and instructive view on topological insulating phases. 
\end{abstract}


\maketitle

\paragraph{Introduction}  

Recently, synthetic magnetic fields \cite{lin2009b} and spin-orbit couplings \cite{lin2011a} have been realized for ultra cold atoms using suitably arranged lasers that couple different internal states~\cite{dalibard2011a}. This opens the path to the experimental investigation of topological phases, such as quantum Hall (QH) states, topological insulators and superconductors, in a clean and highly controllable environment \cite{bloch2008,cooper2008a}. Topological phases currently attract the attention of the scientific community for their remarkable properties, such as dissipationless transport and quantized conductivities \cite{HasanKane2010,qi2011a}. In this context, the recent experimental realization of a staggered magnetic field in a 2D optical lattice, exploiting laser-induced gauge potentials, constitutes an important step in the field \cite{aidelsburger2011a} (cf. also \cite{Jimenez2012a,struck2012a}). In the near future, large {\it uniform} magnetic flux should be reachable using related proposals \cite{jaksch2003a,gerbier2010a,Jimenez2012a,struck2012a}, allowing optical-lattice experiments to explore the Hofstadter model \cite{Hofstadter1976}. The latter is the simplest tight-binding lattice model exhibiting topological transport properties: well-separated energy bands are associated to non-trivial topological invariants, the Chern numbers, leading to a quantized Hall conductivity when the Fermi energy is located in the bulk gaps \cite{thouless1982a,Kohmoto1989a}. These transport properties are directly related to the existence of chiral edge states: while bulk excitations remain inert, these gapless states carry current along the edge of the system. According to the bulk-edge correspondence \cite{Hatsugai1993,Qi2006}, the Chern numbers characterizing the bulk bands determine the number of edge excitations and their chirality, which protects the edge transport against  small  perturbations.

In view of the experimental progress \cite{aidelsburger2011a,Jimenez2012a,struck2012a},  an important issue is to identify observables that provide unambiguous signatures of topological phases in a cold-atom framework \cite{umucalilar2008a,Goldman2010a,StanescuEA2010,Liu2010a,alba2011a,Zhao2011,Kraus2012,Price2012,Buchhold2012,Dellabetta2012}.  This is a crucial topic from an experimental point of view: Measuring the Hall conductivity is more difficult than for solid-state systems due to the absence of particle reservoirs coupled to the system. Several proposals exist for measuring the bulk topological invariants, {\it e.g.} based on spin-resolved time-of-flight \cite{alba2011a} or density measurements \cite{umucalilar2008a,Zhao2011}. On the other hand, a direct detection of the edge states, associated with a proof of their chiral nature \cite{StanescuEA2010,Liu2010a}, would witness the non-trivial topological order associated to cold-atom QH insulators. In this Letter, we propose a realistic and efficient scheme to probe the chiral edge states of the \emph{Hofstadter optical lattice}, which we believe could be extended to any ultracold-atom setup emulating 2D topological phases. 

\paragraph{The model} 

We study a two-dimensional fermionic gas confined in a square optical lattice and subjected to a uniform synthetic magnetic field $\bs B=B \hat 1_z$ \cite{jaksch2003a,gerbier2010a}. The Hamiltonian is taken to be
\begin{align}
\hat H_0=& -J \sum_{m,n} \hat c^{\dagger}_{m+1,n} \hat c_{m,n} + e^{i 2 \pi \Phi m}  \hat c^{\dagger}_{m,n+1} \hat c_{m,n} + \text{h.c.} \notag \\
&+ \sum_{m,n} V_{\text{conf}}(r) \, \hat c^{\dagger}_{m,n} \hat c_{m,n},\label{ham}
\end{align}
with $\hat c_{m,n}$ the field operator defined at lattice site $(x=ma,y=na)$, $m,n \in \mathbb{Z}$, $J$ the tunneling amplitude, and $V_{\text{conf}}(r)$ a cylindrically symmetric confining potential. Eq.~(\ref{ham}) describes non-interacting fermions on a lattice in the tight-binding regime, subjected to a vector potential $\bs{A}=(0, B x,0)$ corresponding to $\Phi$ magnetic flux quanta per unit cell \cite{Hofstadter1976}.  

\begin{figure}[h!]
	\centering
	\includegraphics[width=1\columnwidth]{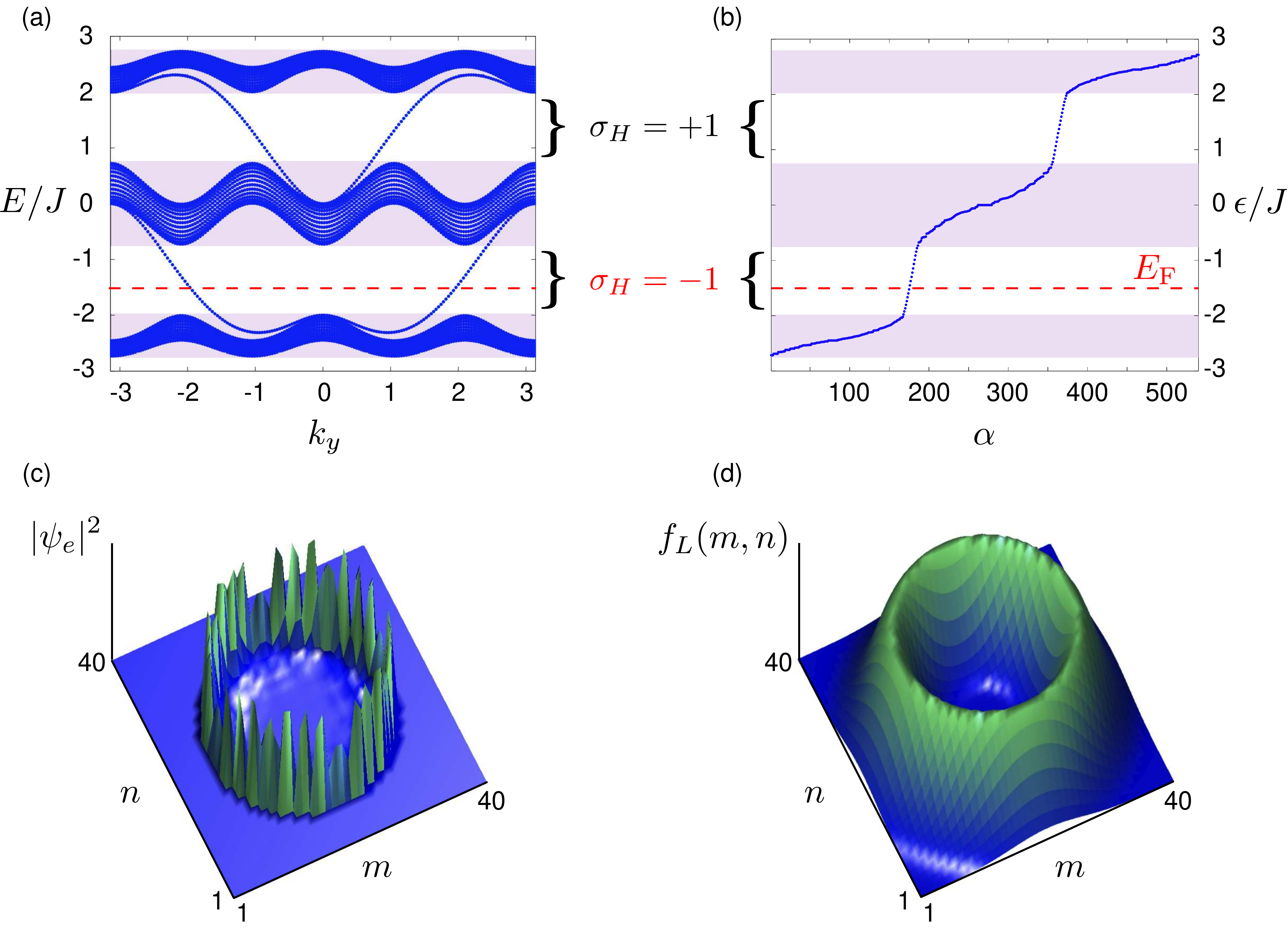}
	\caption{\label{figure1} (Color online) (a) Single-particle energy spectrum $E=E(k_y)$ for the cylindrical lattice subjected to a uniform flux $\Phi \!=\!1/3$, where $k_y$ is the momentum along the closed direction. The bulk states are found in the purple bands and the Hall conductivity $\sigma_{\text{H}} \!=\! \pm 1 $ is indicated for $E_{\text{F}} \! \approx \! \pm 1.5 J$. (b) Discrete energy spectrum for a circular infinite wall with $r_{\text{edge}} \!=\!13 a$. The Fermi energy  $E_{\text{F}} \!=\! -1.5 J$ is represented by a red dotted line and $\alpha$ labels the single-particle states, \emph{i.e.} $\hat{h}_0 \vert \psi_{\alpha} \rangle \!=\! \epsilon_{\alpha} \vert \psi_{\alpha} \rangle$, where $\hat{h}_0$ is the single-particle Hamiltonian.  (c) The amplitude $\vert \psi_e (m,n) \vert^2$ corresponding to the edge state at $\epsilon_e \approx -1.5 J\!=\! E_{\text{F}}$ in (b). \!(d) The probe shape $f_L (m,n)\!=\!f_L (r)$ used to detect the edge state in (c).}
\end{figure}

It is instructive to first discuss  the spectrum $E(k_y)$ obtained by solving the Hamiltonian \eqref{ham} on an abstract cylindrical geometry, without the potential $V_{\text{conf}}(r)$ ($k_y$ denotes the momentum along the closed direction $y$). For $\Phi = p/q$, where $p,q \in \mathbb{Z}$, the spectrum can be partitioned in terms of \emph{bulk states} and \emph{topological edge states}, as shown in Fig.~\ref{figure1}(a). Bulk states exist in the absence of boundary and form $q$ well-separated subbands \cite{Hofstadter1976}. Because of its finite boundaries along $x$, this system also features topological edge states, which propagate along the edges of the cylinder \cite{Hatsugai1993}. They are located within the bulk energy gaps, with a quasi-linear local dispersion relation, $E/\hbar \approx v_{e} k_y$, where $v_e$ is the group velocity. In the spectrum represented in Fig. \ref{figure1} (a), the two bulk gaps are associated to the quantized Hall conductivities $\sigma_H=\pm 1$ (in units of the conductivity quantum), as they host a single edge-state branch per edge, with opposite chirality $\text{sign}(v_e)$. In the following, we set $\Phi=1/3$ and choose  a Fermi energy $E_F=-1.5~J$ located within the first bulk gap, so that the fermionic gas forms a QH insulator with central density $n=1/3a^2$.

The clear partition of the single-particle spectrum into bulk states and topological edge states still holds in the experimental planar geometry and in the presence of the confining potential $V_{\text{conf}}(r)$, taken here of the form $V_{\text{conf}}(r) =J  \left(r/r_{\rm edge} \right)^\gamma $. We illustrate the bulk-edge partition in Fig.~\ref{figure1} (b), which shows the discrete energy spectrum $\epsilon_{\alpha}$ for an infinite circular wall ($\gamma \rightarrow \infty$), with $r_{\text{edge}}=13 a$: Similarly to the cylindrical case, we obtain bulk states within the energy bands and edge states (illustrated in Fig.~\ref{figure1}(c)) within the bulk gaps. We stress that the number of edge excitation branches (relative to the number of physical edges) and their chirality do not depend on the particular geometry, as they are dictated by topological invariants associated to the bulk (see \cite{Hatsugai1993,Qi2006} and Supplementary Material). We verified that the above picture remains valid for finite confinements $\gamma \geq 2$: In agreement with their topological nature, edge states survive for a sufficiently weak confining potential that does not radically perturb the band structure.

\paragraph{Angular Momentum Spectroscopy :}
%

Our aim is to design an experimental probe yielding a clear signature from the topological states. We note that a finite confining potential $V_{\text{conf}}(r)$ generally leads to non-topological edge states \cite{StanescuEA2010,hooley2004a,ott2004a}. However, QH edge states have a crucial property, their chirality (i.e. the sign of their angular velocity), that allows to distinguish them from the bulk and non-topological states by using a probe sensitive to angular momentum. Bragg spectroscopy, a form of momentum-sensitive light scattering, can provide such a probe \cite{Liu2010a,StanescuEA2010}, as we now explain. In its usual implementation  \cite{stenger1999b,steinhauer2002a}, Bragg spectroscopy probes the linear momentum distribution. Here, we propose {\bf (a)} to use a spatial mode carrying angular momentum in order to probe the angular momentum distribution, and {\bf (b)} to shape the probing lasers to maximize the probability to excite an edge state. We consider two lasers denoted $1,2$ in high-order Laguerre-Gauss modes with optical angular momenta $l_{1,2}$, corresponding to the electric fields $E_{1,2}(r) = \sqrt{I_{1,2}} f_{l_1,l_2} (r) \exp (-i l_{1,2} \theta - i \omega_{1,2} t)$, where the radial mode functions are $f_{l}(r)\propto (r/r_0)^{\vert l \vert} e^{-r^2/2 r_0^2}$, and $(r, \theta)$ are polar coordinates. We assume that the beams are set off-resonance from a neighboring atomic transition, so that spontaneous emission can be neglected. This leads  to a scattering Hamiltonian
\begin{align}
\hat H_{\text{Bragg}}(t) &=  \hbar \Omega \sum_{\alpha \beta} \bigl (  I_{\alpha \beta}^q e^{-i \omega_L t} +   I_{\alpha \beta}^{-q} e^{i \omega_L t} \bigr ) \hat c_{\alpha}^{\dagger} \hat c_{\beta},\\
I_{\alpha \beta}^q &=  \frac{1}{2} \int \text{d}\bs{x}\, \psi^{*}_{\alpha} (\bs{x})\psi_{\beta} (\bs{x}) f_L(r) e^{i q \theta}. \label{integrale}
\end{align}
Here, the index $q\!=\!l_2 - l_1$ represents the amount of angular momentum transferred by the probe (in units of $\hbar$),  $\hbar\omega_L\!=\!\hbar(\omega_1 - \omega_2)$ is the energy transfer,  $\Omega$ is the Rabi frequency characterizing the strength of the atom-light coupling, and the probe profile is   $f_{L}(r)\!=\!(r/r_0)^L e^{-r^2/r_0^2}/\mathcal{N}_L$, with $L\!=\!\vert l_1 \vert+\vert l_2 \vert$ (cf. Fig. \ref{figure1} (d)). The operator $\hat c_{\alpha}^{\dagger}$ creates a particle in the eigenstate $\vert \psi_{\alpha} \rangle$ of the unperturbed single-particle Hamiltonian, i.e. $\hat{h}_0 \vert \psi_{\alpha} \rangle \!=\! \epsilon_{\alpha} \vert \psi_{\alpha} \rangle$. 

\begin{figure}[h!]
	\centering
	\includegraphics[width=1\columnwidth]{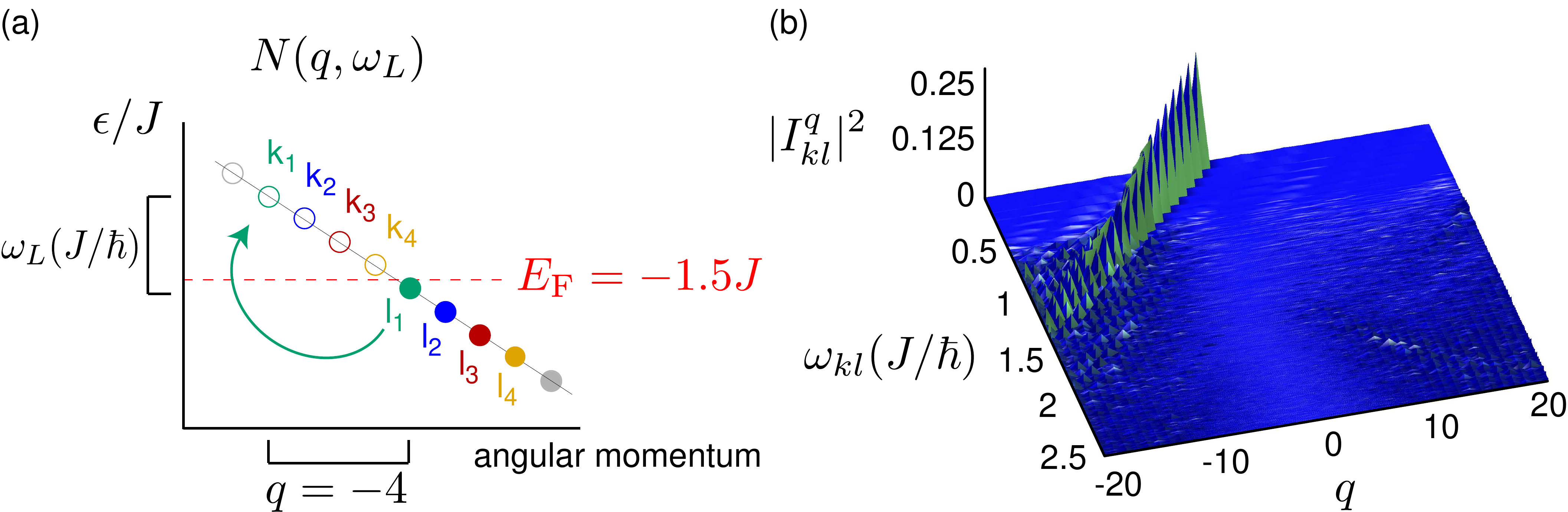}
	\caption{\label{figure2} (Color online) (a) Sketch of the single-particle energy spectrum $\epsilon_{\alpha}$ and the transitions $\vert \psi_{l} \rangle \rightarrow \vert \psi_{k} \rangle$ probed by $N(q, \omega_L)$, for $\omega_L \approx \omega_{kl} \ll J/\hbar$. (b) The amplitude $\vert I_{kl}^q\vert ^2$, as a function of the probe parameter $q$ and excitation frequency $\omega_{kl}$, for $\Phi=1/3$, $E_{\text{F}}=-1.5J$, $L=13$ and $r_0=5.1 a$. The confining potential is infinite ($\gamma = \infty$) and $r_{\text{edge}}=13 a$.} 
\end{figure}

Solving the time-dependent problem $\hat H_0+\hat H_{\text{Bragg}}(t)$ to first order, we write the many-body wave function as 
$\vert \Psi (t) \rangle \approx  b_0 (t) \vert 0 \rangle + \sum_{(k,l)} b_{kl} (t) e^{-i \omega_{kl} t}  \vert k l \rangle$,
where $\vert 0 \rangle=\prod_{\nu \le E_{\text{F}}} \hat c_{\nu}^{\dagger} \vert \emptyset \rangle$ denotes the groundstate \footnote{We will work at zero temperature, but expect that the features discussed here should survive for $T \leq \Delta$, with $\Delta \simeq J$ the energy gap to the next empty band.},  and where $\vert kl \rangle=\hat c_{k}^{\dagger} \hat c_{l} \vert 0 \rangle$ denotes an excited state with a single fermionic excitation ($k > E_{\text{F}},l \le E_{\text{F}}$) with energy 
$\hbar \omega_{kl}=(\epsilon_k -\epsilon_l ) >0$.  With the initial conditions $(b_0(0)\!=\!1, b_{kl}(0)\!=\!0)$, the number of scattered particles (hereafter referred to as \emph{excitation fraction}) is given by
\begin{equation}
N(q,\omega_L) = \sum_{k,l} \vert b_{kl} (t) \vert^2 = \Gamma_{\rm sc} t,
\end{equation}
where the scattering rate $\Gamma_{\rm sc}$ is given by the Fermi golden rule (cf. Supplementary Material for details)
\begin{equation}
\Gamma_{\rm sc}  = 2 \pi \Omega^2  \sum_{k > E_{\text{F}},l \le E_{\text{F}}} \vert I_{kl}^q \vert^2 \delta^{(t)} (\omega_{kl} - \omega_L) ,\label{nfraction}
\end{equation}
with $\pi t \delta^{(t)}(\omega)\!=\!(\sin(\omega t)/\omega)^2$. When the Fermi energy $E_{\text{F}}$ is set within a bulk gap, and for small intensities $\hbar \Omega \ll J $, the excitation fraction $N(q,\omega_L) $ probes the dispersion relation $\epsilon_e=\epsilon_{\text{e}} (M)$ associated to the gapless edge states $\vert \psi_e \rangle$ that lie within this gap, where $M$ is a quantum number analogous to angular momentum (see Fig.~\ref{figure2} (a)). For an optimized probe shape  (obtained for $r_0 \approx r_{\text{edge}}/ \sqrt{\vert L \vert/2}$), this can be deduced from the behavior of the overlap integrals $I_{kl}^q$ defined in Eq.~\ref{integrale}. They are represented in the $\omega_{kl}-q$ plane in Fig.~\ref{figure2}(b) for $\gamma= \infty$. At low frequencies $\omega_{kl} \ll J / \hbar$, we find a continuous alignment of resonance peaks $\omega_{kl}^{\text{res}}\approx \dot{\theta}_{\text{e}} q$. This reflects the linear dispersion relation $\epsilon/\hbar \approx \dot{\theta}_{\text{e}} M$ in the vicinity of the Fermi energy, and provides the angular velocity $\dot{\theta}_{\text{e}} \approx -0.07 J/\hbar$ and the chirality (i.e. $\text{sign}(\dot{\theta}_{\text{e}})$) characterizing the edge states in the lowest bulk gap. We find that this result is in perfect agreement with a direct evaluation of the angular velocity \footnote{The angular velocity can be directly evaluated through $\dot{\theta}_{\text{e}}=(i/\hbar) \sum_{m,n} \psi_e^{*} (m,n) [\hat{h}_0, \hat \theta] \psi_e (m,n)$, where $\hat \theta$ is the polar angle operator.}. We emphasize that the edge states velocity highly depends on the boundary produced by the confinement: $\dot{\theta}_{\text{e}}$ significantly decreases as the potential $V_{\text{conf}}(r)$ is smoothened  (\emph{e.g.} $\dot{\theta}_{\text{e}}\approx-0.02 J/\hbar$ for $\gamma=10$, $\dot{\theta}_{\text{e}}\approx -0.01 J/\hbar$ for $\gamma=2$). The absence of substantial response for $q>0$  in Fig.~\ref{figure2} (b) clearly proves that our setup is effectively sensitive to the edge state chirality. Naturally, the signal obtained by setting the Fermi energy in the second bulk gap, or by reversing the sign of the magnetic flux $\Phi \rightarrow - \Phi$, would probe the \emph{opposite} chirality. 

\begin{figure}[h!]
	\centering
	\includegraphics[width=1\columnwidth]{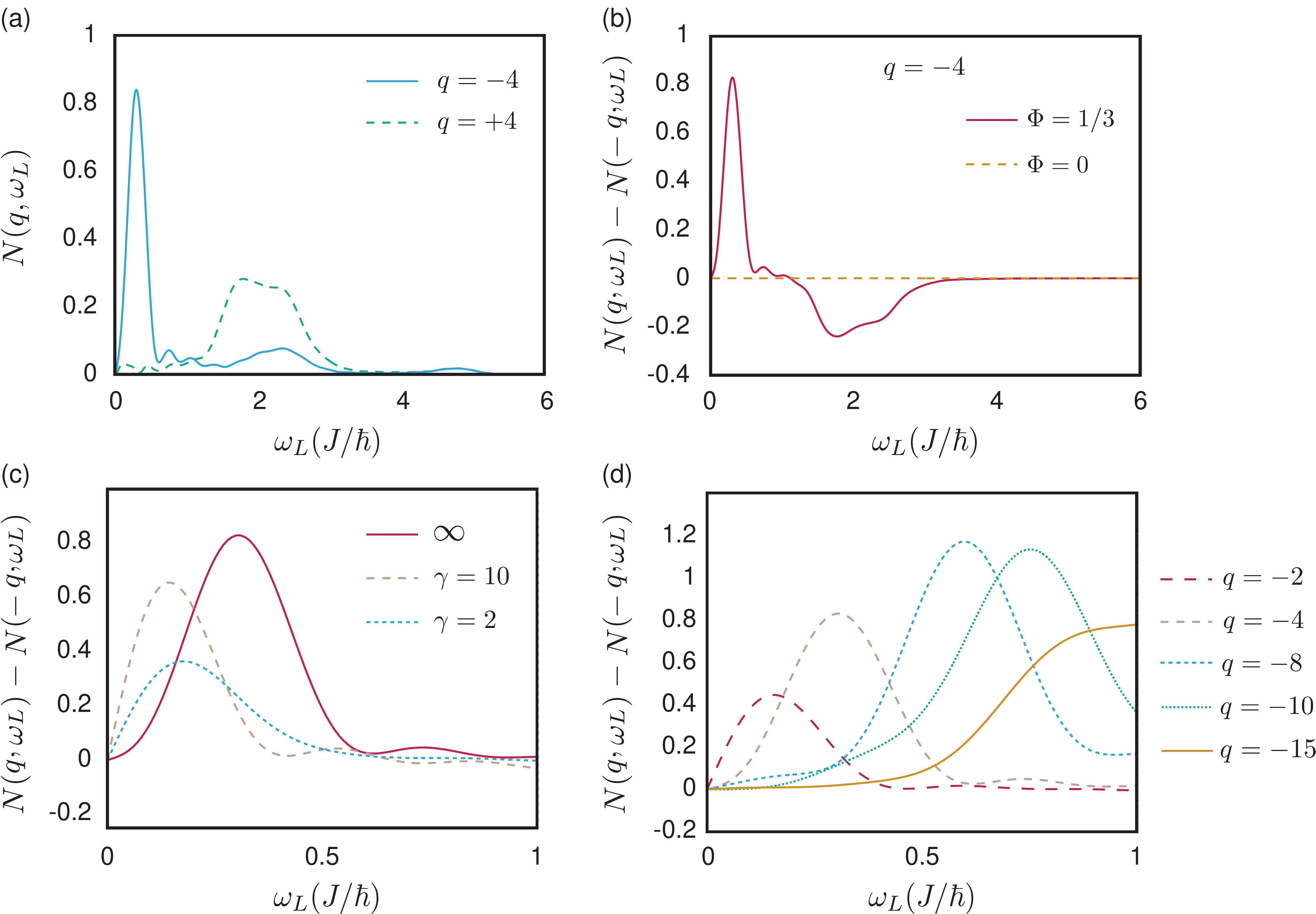}
	\caption{\label{figure3} (a) Excitation fraction $N(q,\omega_L)$ versus probe frequency, for an angular momentum transfer $q\!=\!\pm 4$.  (b) $N(q,\omega_L)\!-\!N(-q,\omega_L)$ for $\Phi\!=\!0$ and $\Phi\!=\!1/3$. (c) $N(q,\omega_L)\!-\!N(-q,\omega_L)$ for several shapes of the trapping potential $V_{\text{conf}}(r) \!=\! J  \left(r/r_{\rm edge} \right)^\gamma $ and $q \!=\! -4$. (d) Increasing then broadening of $N(q,\omega_L)\!-\!N(-q,\omega_L)$, for increasing $\vert q \vert$. In all the figures $\Omega\!=\!0.05 J/\hbar$, $t\!=\!20 \hbar/J$, $L\!=\!13$, $r_0\!=\!5.1 a$, $r_{\text{edge}}\!=\!13 a$, $\Phi\!=\!1/3$, $E_{\text{F}}\!=\!-1.5 J$ and $\gamma=\infty$ (except in (c)).} 
\end{figure}

We obtain the excitation fraction $N(q, \omega_L)$ at finite times through a direct numerical resolution of the Schr\"odinger equation \footnote{We use excitation times of several $\hbar/J$, typically, which seem experimentally realistic. This is long enough to resolve the edge-edge resonance but still too short to  neglect the broadening due to the finite pulse time  (cf. Supplementary Material).}. A typical result is presented in Fig.~\ref{figure3}(a), for $q=\pm 4$, emphasizing the three distinct regimes of light scattering: ``edge-edge", ``bulk-edge" and ``bulk-bulk". The ``edge-edge" regime  corresponds to transitions solely performed between the edge states close to $E_{\text{F}}$: A sharp resonance peak is visible at $\omega_{L}^{\text{res}}\approx \dot{\theta}_{\text{e}} q\approx 0.3 J/\hbar$  for $q=-4$, and stems from four transitions between edge states, as sketched in Fig.~\ref{figure2}(a). Then, at higher frequencies, $\omega_L \approx J/\hbar$, small peaks witness allowed transitions between the lowest bulk band and the edge states located above $E_{\text{F}}$. Finally, for $\omega_L \approx 2 J/\hbar$, many transitions between the two neighboring bulk bands lead to a wide and flat signal. This bulk-bulk response is significant for both $q=\pm 4$, as a consequence of the large density of excited states in this frequency range. We stress that a well focused probe allows to significantly reduce any signal of the bulk. In the following, we consider the quantity $N(q, \omega_L) -N(-q, \omega_L)$, which is zero for a system with time-reversal symmetry (cf. Fig.~\ref{figure3}(b)). We have repeated the calculations for several potential shapes,  finding no qualitative change  (cf. Fig.~\ref{figure3}(c)). Although it is advantageous to use a steep confining potential,  the signal from the edge states is robust, even in a harmonic trap ($\gamma=2$).   

The dispersion relation being almost linear close to $E_{\text{F}}$, the number of allowed transitions $\vert \psi_{l} \rangle \rightarrow \vert \psi_{k} \rangle$ scales with the probe parameter $q$ in the ``edge-edge" regime (cf. Fig.~\ref{figure2}(a)). Thus,  one observes an increase of the peaks for increasing values of $\vert q \vert$ (cf. Fig. ~\ref{figure3}(d)). We stress that this progression only occurs in the ``edge-edge" regime, namely when $q$ is chosen such that $\hbar \omega_{L}^{\text{res}} \approx \hbar \dot{\theta}_{\text{e}} q$ is smaller than the energy difference between $E_{\text{F}}$ and the closest bulk band. In the case illustrated in Fig. ~\ref{figure3}(d), the ``edge-edge" regime is delimited by $\vert q_{\text{e-e}} \vert \lesssim 7$. Beyond $\vert q_{\text{e-e}} \vert$, the resonance peak enters the ``bulk-edge" regime: The excitation fraction $N(q, \omega_L)$ broadens, $N(-q, \omega_L)$ is no longer negligible, and the linear dispersion relation is no longer probed. We thus conclude that a moderate value (here $\vert q \vert \sim4$) is preferable to keep a narrow peak, well separated from the broader ``edge-bulk" signal.

\paragraph{Imaging the edge states}

With respect to experimental detection, the Bragg scheme described above presents a major drawback. The associated signatures in the spatial or momentum densities are small perturbations on top of the strong ``background" of unperturbed atoms. For a circular system with Fermi radius $R_F$, one can expect about $N_{\rm edge}\sim \Delta/\hbar \dot{\theta}_e\approx R_F \Delta / \hbar \vert v_e \vert$ edge states, with $\Delta$ the bulk energy gap  and $v_e$ the group velocity (cf. above). Using the parameters from Fig.~\ref{figure3}(a), one finds $N_{\rm edge}\sim 20$ while the total number of atoms in the calculation is $N\sim 200$. Scaling to more realistic numbers for an experiment ($N\sim 10^4$), and noting that the number of scattered atoms is at most a fraction of $N_{\rm edge}$, we conclude that one should be able to detect a few tens of atoms at best on top of the signal coming from $\sim 10^4$ unperturbed ones: This is a significant experimental challenge with present-day technology. One possibility to avoid this difficulty is to use an alternate detection scheme, where the probe also changes the atomic internal state. The probe signal can then be measured against a dark background (without unperturbed atoms), which allows powerful imaging methods to be used (\emph{e.g.} large aperture microscopy, as recently demonstrated for quantum gases in optical lattices \cite{bakr2009a,sherson2009a}) . 

\begin{figure}[h!]
	\centering
	\includegraphics[width=1\columnwidth]{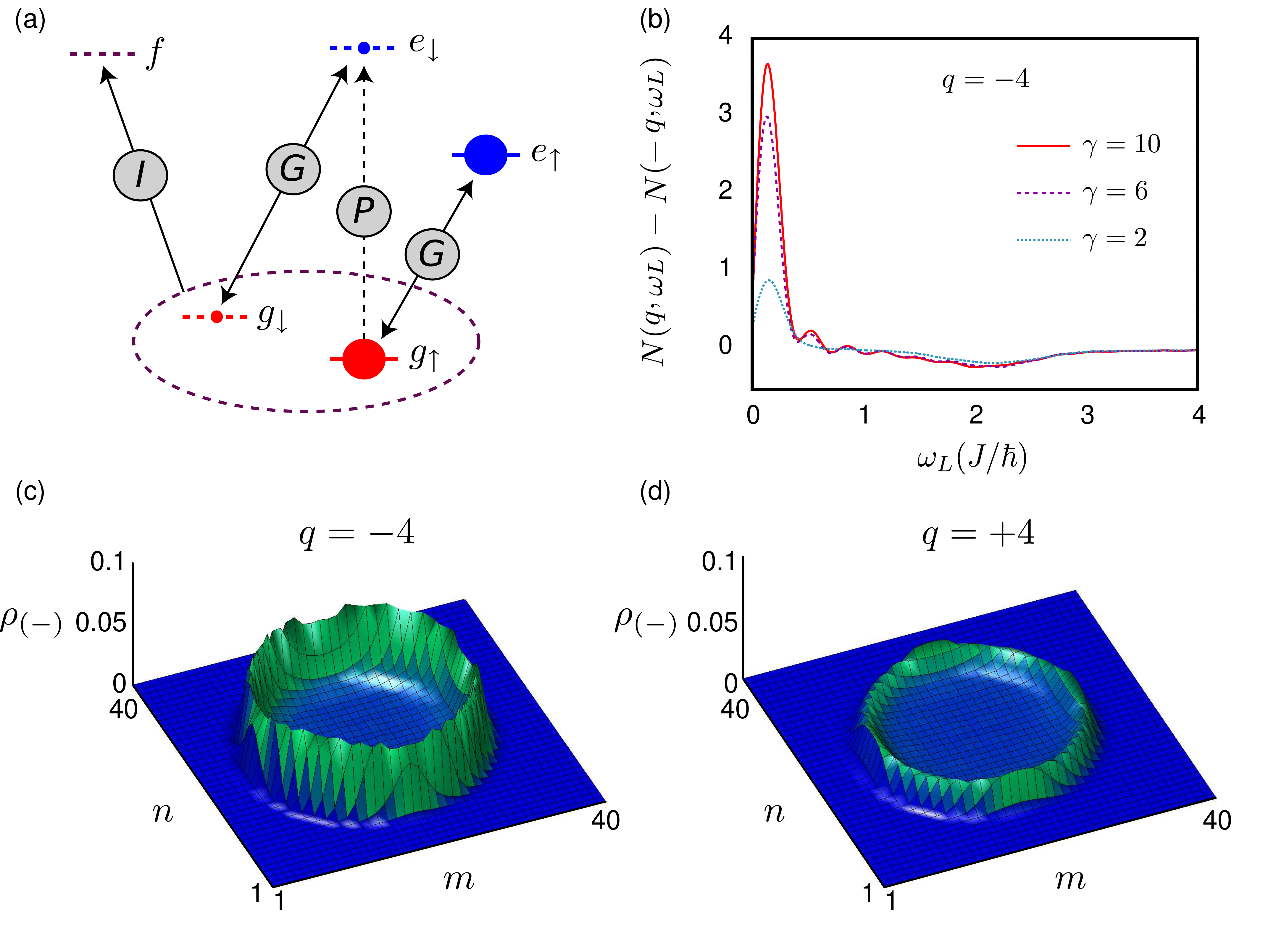}
	\caption{\label{figure4} (a) Detection scheme for a two-electron atom, where $g$ is  the electronic ground state, $e$ a metastable excited state and $f$ a second excited state with short radiative lifetime, suitable for imaging. The coupling lasers $G$ induce an artificial gauge field and the probe laser $P$ allows to detect edge states (cf. text). The circles indicate relative populations in each state after the probe pulse. (b) Excitation fraction versus probe frequency, for several potential shapes. (c)-(d) Density $\rho_{(-)} (m,n)$ for $q=\pm 4$, $\gamma=10$ and $\omega_{L} = 0.14 J/\hbar$. The probe and system parameters in (b)-(d) are the same as in Fig.\ref{figure3}.} 
\end{figure}

We present in Fig.~\ref{figure4}(a) a possible implementation suitable for two-electron atoms with ultra-narrow optical transitions, inspired by the electron shelving method. In the following, we use  the particular case of  $^{171}$Yb atoms  for illustration (see also Supplementary Material).  The ground $g$ and excited $e$ states have nuclear spin $I=1/2$, leading to ground $\{g_\downarrow,g_\uparrow \}$ and excited $\{e_\downarrow,e_\uparrow \}$ manifolds. The Zeeman degeneracies are split by a real and relatively strong magnetic field $B\sim 100~$G. The states $g_\uparrow$ and $e_\uparrow$ are initially populated, as laser coupling between these two states is used to generate the artificial gauge field leading to Eq.~(\ref{ham}) (cf. methods described in \cite{jaksch2003a,gerbier2010a}). In order to probe the $(+)=\{g_\uparrow,e_\uparrow\}$ system, one introduces a weaker additional laser, coupling $g_\uparrow \rightarrow e_\downarrow$. A crucial point is to ensure that topological edge states have the same structure in the initial and final states. To this end, the initially unpopulated  states $(-)=\{g_\downarrow,e_\downarrow\}$ are also coupled by a laser generating the same gauge field as for $(+)$. After the probe pulse, atoms in the $g$ manifold are dispatched (possibly detected) using an auxiliary imaging transition $g \rightarrow f$. Crucially, atoms in the $e$ manifold are not in resonance with the imaging light and are therefore unaffected. The $e_\downarrow$ atoms are subsequently brought down to the $g$ manifold using, {\it e.g.}, adiabatic-passage techniques, and they are finally detected, without stray contributions from unperturbed atoms in other internal states.

To analyze the effect of this probe, we consider a simplified level scheme with two internal states only, denoted by the indices $(\pm)$.  We suppose that only the $(+)$ sector is initially populated. The spatial profile of the coupling laser is similar to the one used for Bragg excitations, but now the Pauli principle does not  restrict the available final states, since the state $(-)$ is initially unoccupied. The calculation proceeds as before, with initial $\vert 0 \rangle= \prod_{\epsilon_{\nu} \leq E_F} \hat{c}_{\nu,+}^\dagger \vert \emptyset \rangle$ and final $\vert kl \rangle= \hat{c}_{k,-}^\dagger \hat{c}_{l,+}\vert 0 \rangle$ states.
We write the coupling to the probe as
\begin{align}
&\hat H_{\text{Shelving}}(t)= \hbar \Omega \sum_{\alpha \beta} I_{\alpha \beta}^q \hat c_{\alpha,-}^{\dagger} \hat c_{\beta,+} e^{-i \omega_L t} +  \text{h.c.},\label{Shelving}
\end{align}
where the operator $\hat c_{\alpha, (\pm)}^{\dagger}$ creates a particle of the $(\pm)$ sector in the eigenstate $\vert \psi_{\alpha} \rangle$, and where $I_{\alpha \beta}^q$ has the same definition as in Eq. \eqref{integrale}, since $\hat h _{+}= \hat h _{-}=\hat h _{0}$. We also suppose that $\hbar \Omega \ll J$ to neglect higher order excitations. 

The excitation fraction $N(q,\omega_L)-N(-q,\omega_L)$ is represented in Fig. \ref{figure4}(b), showing a clear resonance peak at low frequencies $\hbar  \omega_{L}  \ll J$. Interestingly, this result shows that the low-energy regime is still governed by the chiral edge states located in the bulk gap, although transitions are now allowed for all the states below $E_{\text{F}}$, including the bulk states (see Supplementary Material).
Indeed, the signal $N(q,\omega_L)-N(-q,\omega_L)$ remains small and flat in the ``edge-bulk" region, while the chiral ``edge-edge" peak stands even clearer than in the Bragg case (since more ``edge-edge" transitions are allowed between states of same chirality). By setting the probe parameters $(q, \omega_L)$ close to a resonance peak, one can now populate edge states into the $(-)$ sector and directly visualize them using state-selective imaging. The corresponding density $\rho_{(-)} (m,n)=\langle \Psi (t) \vert \hat n_{(-)} (m,n)\vert \Psi (t) \rangle$ is illustrated in Figs. \ref{figure4}(c)-(d) for $q=\pm 4$. The clear difference between the two images, obtained with different signs of $q$ but otherwise identical setups, is a direct proof of the chiral nature of the edge excitations populated by the probe. Thus, we have demonstrated that our state-dependent probe \eqref{Shelving}, combined with \emph{in situ} imaging, provides an efficient method to identify topological edge states in a cold-atom experiment. We believe that this method could be generalized to any cold-atom setup subjected to synthetic gauge potentials that emulate 2D topological phases, even in the presence of interactions or disorder.

We thank the F.R.S-F.N.R.S, DARPA (Optical lattice emulator project), the Emergences program (Ville de Paris and UPMC)  and ERC (Manybo Starting Grant) for financial support. We are grateful to J. Dalibard, S. Nascimb\`ene, F. Ch\' evy, C. Morais Smith, P. {\"O}hberg, W. Hofstetter, M. Buchhold and D. Cocks for inspiring discussions and comments. 
%


\appendix


\section{The bulk-edge correspondence and the circular geometry}

In this work, we consider the Hofstadter optical lattice described by the Hamiltonian in Eq. (1). In an optical-lattice experiment, the atoms are confined by a cylindrically symmetric confining potential, which we take of the form
\begin{equation}
V_{\text{conf}}(r) =J  \left(r/r_{\rm edge} \right)^\gamma , \label{confining}
\end{equation}
where the parameters $J$, $r_{\rm edge}$ and $\gamma$ are defined in the main text. We are interested in the detection of chiral edge states, which propagate along the circular edge of this system. In the solid-state framework, these chiral states are responsible for the quantum Hall effect \cite{HasanKane2010,qi2011a,Hatsugai1993}. In this appendix, we recall how these \emph{topological} edge states are indeed related to the concept of topological invariants, which guarantee their robustness against small external perturbations. 

The fundamental concept which relates the chiral edge states illustrated in Fig. 1 (c) to topological invariants is the so-called \emph{bulk-edge correspondence} \cite{Hatsugai1993,Qi2006}. This theorem stipulates that when specific topological invariants associated to the \emph{bulk} bands are non-zero, the presence of \emph{edge} states at the boundaries is guaranteed and that their chirality (i.e. their orientation of propagation along the edge) is fixed. Moreover, the edge states energies are located within the bulk gaps, as illustrated in Fig. 1 (a). Here, the topological invariants are defined on the first Brillouin zone, or equivalently on an abstract two-dimensional torus $\mathbb{T}^2$, as a result of periodic boundary conditions applied to both spatial directions (more precisely, the topological invariants are defined on a principle fibre bundle $P(\text{U(1)},\mathbb{T}^2)$, which is based on the two-torus $\mathbb{T}^2$). These topological indices are the Chern numbers $N_{\nu}$, which are associated to each bulk band $E_{\nu} (\bs k)$ through the  Thouless-Kohmoto-Nightingale-Nijs expression (TKNN) \cite{thouless1982a}
\begin{equation}
N_{\nu}=  \frac{i}{2 \pi} \int_{\mathbb{T}^2} \langle \partial_{k_x} u_{\nu}(\bs k) \vert \partial_{k_y} u_{\nu}(\bs k) \rangle - (k_x \leftrightarrow k_y) \txt{d} \bs{k},\label{chern}
\end{equation}
where $\vert u_{\nu}(\bs k) \rangle$  is the single-particle eigenstate of the Hamiltonian with energy $E_{\nu} (\bs k)$, and $\bs k=(k_x,k_y) \in \mathbb{T}^2$ is the quasi-momentum. When the Fermi energy $E_{\text{F}}$ is exactly located in a bulk gap, the Hall conductivity is directly related to the Chern numbers,
\begin{equation} 
\sigma_H= \sum_{E_{\nu}<E_{\text{F}}} N_{\nu}, \label{Hall}
\end{equation} 
which can be directly derived from the Kubo formalism  \cite{thouless1982a,Kohmoto1989a}. Here the conductivity is expressed in units of the conductivity quantum. 

For the specific case studied in this work, where we set the magnetic flux $\Phi=1/3$, the bulk energy spectrum splits into three energy bands, which have the associated Chern numbers $N_1=-1$, $N_2= 2$ and $N_3 = -1$. Therefore, when the Fermi energy lies in the first [resp. second] bulk gap, the Hall conductivity corresponds to $\sigma_H=-1$ [resp. $\sigma_H=+1$], as illustrated in Fig. 1 (a)-(b). These results, which are derived from the toroidal geometry, are summarized in Table \ref{cherntab}.

The bulk-edge correspondence dictates the following result: if we solve the same model (1) on an open geometry (i.e. a system with boundaries), gapless edge states will appear in the bulk gaps. Moreover the number of edge-state branches is given by the modulus of the Hall conductivity in \eqref{Hall}, and their chirality by $\text{sign}(\sigma_H)$. This can be easily visualized by solving the model (1) on an abstract cylinder (with $V_{\text{conf}}=0$). The corresponding energy spectrum $E (k_y)$ is illustrated in Fig. 1 (a) and is described in the main text. Since the cylinder has two physical edges, we note that the first bulk gap in Fig. 1 (a) hosts two edge-state branches with opposite orientations (i.e. one for each physical edge) \cite{Hatsugai1993}. The presence of a single edge excitation (per physical edge) is in agreement with the fact that $\vert \sigma_H \vert=\vert N_1\vert=1$ when the Fermi energy is located in this bulk gap: this is precisely the bulk-edge correspondence, which is summarized in Table \ref{bulkedgetab}. 

\begin{figure}[!]
	\centering
	\includegraphics[width=1.0\columnwidth]{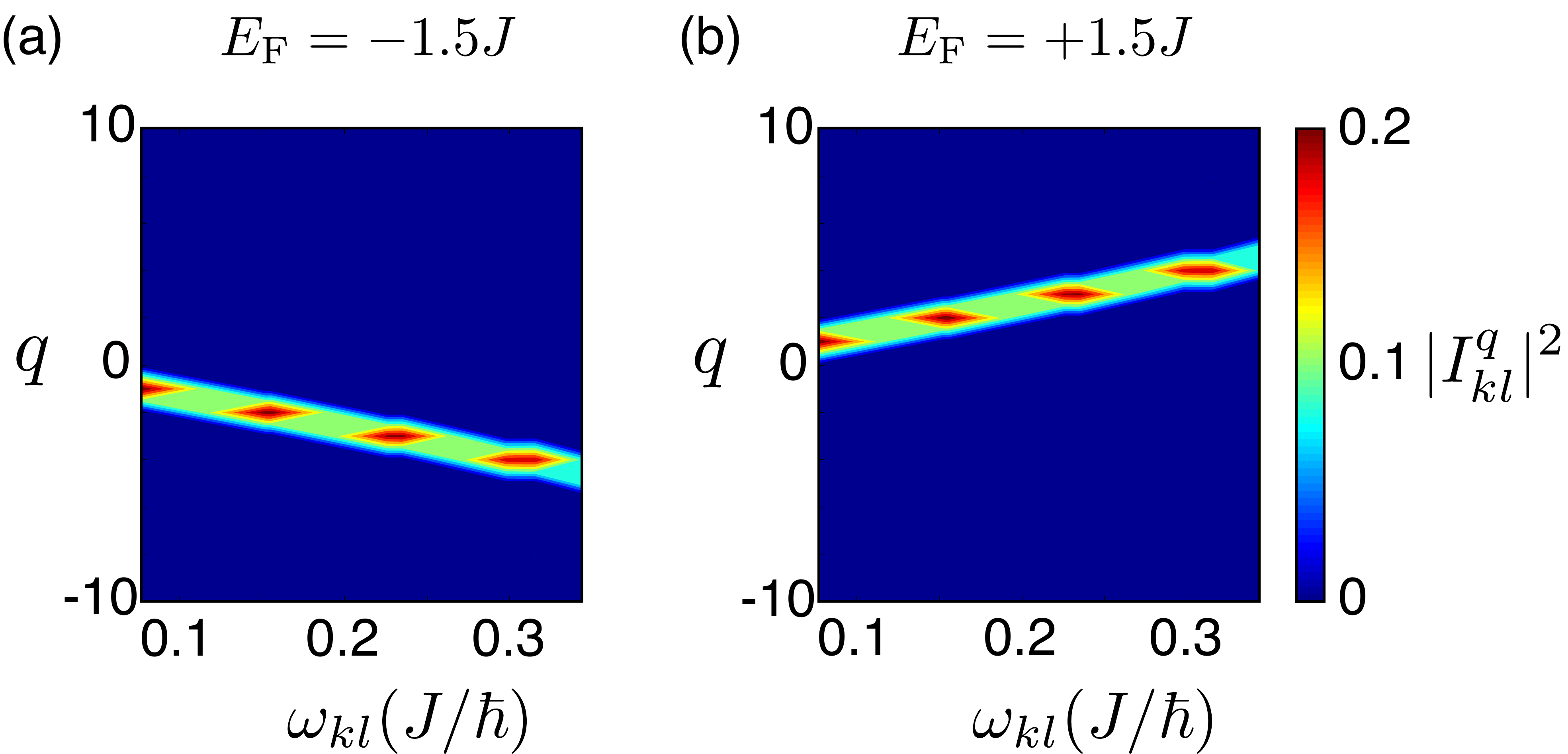}
	\caption{\label{contour} (Color online) The amplitude $\vert I_{kl}^q\vert ^2$, as a function of the probe parameter $q$ and excitation frequency $\omega_{kl}$, for $\Phi=1/3$, $L=13$ and $r_0=5.1 a$. The confining potential is infinite ($\gamma = \infty$) and $r_{\text{edge}}=13 a$. (a) $E_{\text{F}}=-1.5J$ and (b) $E_{\text{F}}=+1.5J$. The angular velocity of the edge states present in the first [resp. second] bulk gap is $\dot{\theta}_{\text{e}} \approx \omega_{kl}^{\text{res}}/q \approx -0.07 J/\hbar$ [resp. $\dot{\theta}_{\text{e}} \approx +0.07 J/\hbar$]. Thus, the two bulk gaps are associated with opposite chiralities, in agreement with the bulk-edge correspondence in Table \ref{cherntab}.} 
\end{figure}

We stress that the number of edge-state branches (per physical edge) is independent of the boundary geometry, as it is given by a sum of topological indices \eqref{Hall} associated to the bulk bands. Therefore, when considering the realistic circular geometry produced by the confining potential $V_{\text{conf}}(r)$ in Eq. \eqref{confining}, one obtains the \emph{same} edge-state structure propagating along the circular edge $r=r_{\text{edge}}$ as the one obtained from the abstract cylinder discussed above: the number of edge-state branches and the chirality deduced from them are identical, as these properties do not depend on the chosen geometry. However, let us comment on the fact that the boundaries do affect the dispersion relations, and thus the angular velocity, of the edge states (cf. main text). Here, the bulk-edge correspondence indicates that the lowest bulk gap in Fig. 1(b), which corresponds to the circular geometry, hosts a single edge-state branch associated to a negative angular velocity, since the corresponding Hall conductivity $\sigma_H=- 1$ is solely governed by the topological expression \eqref{Hall}. Furthermore, the edge-state branch present in the second bulk gap corresponds to the opposite chirality, since $\sigma_H=+1$ when the Fermi energy is in the highest gap. These results have been verified by directly computing the angular velocity of the edge states,  
\begin{align}
\dot{\theta}_{\text{e}}&=(i/\hbar) \sum_{m,n} \psi_e^{*} (m,n) [\hat{h}_0, \hat \theta] \psi_e (m,n) , \nonumber \\
&\! \! \! \! \! \! \! \!= \frac{J i}{\hbar} \sum_{m,n}   \bigl(  \theta(m+1,n) - \theta (m,n) \bigr ) \psi^*_{e} (m,n) \psi_{e} (m+1,n) \nonumber \\
&\! \! \! \! \! \! \! \!+ \bigl(  \theta(m-1,n) - \theta (m,n) \bigr) \psi^*_{e} (m,n) \psi_{e} (m-1,n)     \nonumber     \\
&\! \! \! \! \! \! \! \!+e^{i 2 \pi \Phi m} \bigl(  \theta(m,n+1) - \theta (m,n) \bigr ) \psi^*_{e} (m,n) \psi_{e} (m,n+1)  \nonumber \\
&\! \! \! \! \! \! \! \!+e^{-i 2 \pi \Phi m} \bigl(  \theta(m,n-1) - \theta (m,n) \bigr ) \psi^*_{e} (m,n) \psi_{e} (m,n-1),\nonumber
\end{align}
where $\theta (m,n)=\text{atan2} \bigl ( (n-N/2),(m-N/2) \bigr )$ and $m,n=1, \dots, N$, and also indirectly through the Bragg signals, as illustrated in this Appendix (cf. Fig. \ref{contour}).

\section{Angular Momentum Spectroscopy}

The interaction between the Bragg lasers and the atoms is described by the Hamiltonian
\begin{align}
&\hat H_{\text{Bragg}}(t)= \hbar \Omega \bigl ( \hat{W}_q e^{-i \omega_L t} +  \hat{W}_{-q} e^{i \omega_L t} \bigr ) , \label{pert} \\
&\hat{W}_q=\sum_{\alpha \beta} I_{\alpha \beta}^q \hat c_{\alpha}^{\dagger} \hat c_{\beta},\\
&I_{\alpha \beta}^q = \frac{1}{2} \int \text{d}\bs{x}\, \psi^{*}_{\alpha} (\bs{x})\psi_{\beta} (\bs{x}) f_L(r) e^{i q \theta}. \label{integ}
\end{align}
Here,  the operator $\hat c_{\alpha}^{\dagger}$ creates a particle in the eigenstate $\vert \psi_{\alpha} \rangle$ of the unperturbed single-particle Hamiltonian, i.e. $\hat{h}_0 \vert \psi_{\alpha} \rangle = \epsilon_{\alpha} \vert \psi_{\alpha} \rangle$ (cf. main text). Solving the time-dependent problem $\hat H_0+\hat H_{\text{Bragg}}(t)$ to first order, we write the many-body wave function as
\begin{align}
\vert \psi (t) \rangle &= b_0 (t) \vert 0 \rangle + \sum_{\mu} b_{\mu} (t) e^{-i E_{\mu} t/\hbar}  \vert \Psi_{\mu} \rangle, \nonumber \\
&\approx b_0 (t) \vert 0 \rangle + \sum_{(k,l)} b_{kl} (t) e^{-i \omega_{kl} t}  \vert k l \rangle,\label{wavedev}
\end{align}
where $\vert 0 \rangle=\prod_{\nu \le E_{\text{F}}} \hat c_{\nu}^{\dagger} \vert \emptyset \rangle$ denotes the groundstate at zero temperature, and 
\begin{equation}
\vert k l \rangle= \vert 1 \dots 1 \underbrace{0}_{l} 1 \dots 1 \underbrace{\vert}_{E_{\text{F}}} 0 \dots 0\underbrace{1}_{k} 0 \dots 0 \rangle, \label{myex}
\end{equation}
where $k > E_{\text{F}},l \le E_{\text{F}}$ and $\omega_{kl}=(\epsilon_k -\epsilon_l )/\hbar >0$. Here
we have restricted the full Hilbert space to the subspace spanned by the ground state and the excited states that are coupled to it to first order in the perturbation \eqref{pert}.

Setting the initial condition $(b_0(0)=1, b_{kl}(0)=0)$, one finds 
\begin{equation}
b_{kl} (t)=- i \Omega  \biggl ( I_{kl}^q S_{kl}^- (\omega_L) e^{i \Delta_{kl}^{-} t }+(I_{lk}^q)^{*} S_{kl}^{+} (\omega_L) e^{i \Delta_{kl}^{+} t } \biggr ), 
\end{equation}
where $S_{kl}^{\pm} (\omega_L)= \sin (\Delta_{kl}^{\pm} t)/\Delta_{kl}^{\pm}$, $\Delta_{kl}^{\pm} =(\omega_{kl} \pm \omega_L)/2$. The number of scattered atoms, or excitation fraction, is then given by 
\begin{align}
&N(q,\omega_L) = \sum_{k,l} \vert b_{kl} (t) \vert^2, \nonumber \\
&= \Omega^2 \sum_{k,l} \vert I_{kl}^q S_{kl}^- (\omega_L) e^{i \Delta_{kl}^{-} t }+(I_{lk}^q)^{*} S_{kl}^{+} (\omega_L) e^{i \Delta_{kl}^{+} t } \vert ^2. \label{finitetime}
\end{align}
In the long-time limit, and neglecting the anti-resonnant term $(\propto e^{i \Delta_{kl}^{+} t })$, this yields the standard Fermi golden rule
\begin{equation}
N(q,\omega_L)  = 2 \pi \Omega^2 t  \sum_{k > E_{\text{F}},l \le E_{\text{F}}} \vert I_{kl}^q \vert^2 \delta^{(t)} (\omega_{kl} - \omega_L) , \label{fermiGR}
\end{equation}
where $\delta^{(t)}(\omega)\!= (1/\pi t)(\sin(\omega t)/\omega)^2 \overset{t \rightarrow \infty}{\xrightarrow{\hspace*{1cm}}}  \delta(\omega)$. The expression \eqref{fermiGR} emphasizes the explicit relation between the excitation fraction $N(q,\omega_L)$ and the rates $\vert I_{kl}^q \vert^2$ presented in Fig. 2 (b). \\

At finite times, it is preferable to evaluate the excitation fraction through a numerical evaluation of the Schr\"odinger equation
\begin{align}
&i \hbar \frac{\text{d} b_{kl} (t)}{\txt{d} t}= \hbar \Omega \sum_{n,m} W_{kl;nm} (t) b_{nm} (t) e^{i (E_{kl}- E_{nm}) t /\hbar} , \label{numeric}
\end{align}
where $W_{kl;nm} (t)\!=\!\langle kl \vert \hat{W} (t) \vert  nm \rangle$ and $\hat{W} (t)= \hat{W}_q e^{-i \omega_L t} +  \hat{W}_{-q} e^{i \omega_L t}$. The many-body wavefunction is still restricted to the first-order subspace but off-resonant terms and deviations from the long-time limit are included. For the reasonable finite times and small Rabi frequencies $\Omega \ll J/\hbar $ used in our calculations (cf. Figs. (3)-(4)), we find that the excitation fraction obtained from a numerical evaluation of Eq. \eqref{numeric} is in perfect agreement with Eq. \eqref{finitetime}. Note that we use excitation times of several $\hbar/J$, typically, which seem experimentally realistic. This is long enough to resolve the edge-edge resonance but still too short to neglect the broadening due to the finite pulse time (cf. Figs. (3)-(4)(b)).

\section{The Shelving Method}

The scattering Hamiltonian considered in the ``shelving method" has the form
\begin{align}
&\hat H_{\text{Shelving}}(t)= \hbar \Omega \bigl ( \hat{W}_q^{\text{sh}} e^{-i \omega_L t} +  \bigl (\hat{W}_{q}^{\text{sh}} \bigr)^{\dagger} e^{i \omega_L t} \bigr ) ,\\
&\hat{W}_q^{\text{sh}}=\sum_{\alpha \beta} I_{\alpha \beta}^q \hat c_{\alpha (-)}^{\dagger} \hat c_{\beta (+)},
\end{align}
where the operator $\hat c_{\alpha (\pm)}^{\dagger}$ creates a particle of the $(\pm)$ sector in the eigenstate $\vert \psi_{\alpha} \rangle$, and where $I_{\alpha \beta}^q$ has the same definition as in Eq. \eqref{integ}, since $\hat h _{(-)}= \hat h _{(+)}=\hat{h}_0$.  

In this scheme, we suppose that only the $(+)$ sector is initially populated, such that the initial and excited states have the following forms
\begin{align}
&\vert 0 \rangle\!=\! \vert 1 \dots 1 \underbrace{\vert}_{E_{\text{F}}} 0 \dots 0 \rangle_{(+)} \vert 0 \dots 0 \rangle_{(-)}, \label{shelve1} \\
&\vert kl \rangle\!=\! \vert 1 \dots 1 \underbrace{0}_{l} 1 \dots 1 \underbrace{\vert}_{E_{\text{F}}} 0 \dots 0 \rangle_{_{(+)}} \vert 0 \dots 0\underbrace{1}_{k} 0 \dots 0 \rangle_{_{(-)}}, \label{shelve2}
\end{align}
where we suppose that $\Omega \ll J/\hbar$ to neglect higher order excitations. We note that $k$ is no longer restricted by the Pauli principle, such that $\omega_{kl}=(\epsilon_k -\epsilon_l )/\hbar$ may now take negative values. We follow the same treatment as for the Bragg scheme and we obtain the excitation fraction as
\begin{equation}
N(q,\omega_L)  = 2 \pi \Omega^2 t  \sum_{l \le E_{\text{F}}} \sum_k \vert I_{kl}^q \vert^2 \delta^{(t)} (\omega_{kl} - \omega_L) , \label{fermiGR2}
\end{equation}
which differs from Eq. \eqref{fermiGR} by the fact that the final states $k$ are now unrestricted.  However, we stress that the sum over the initial states, $\sum_{l \le E_{\text{F}}}$ in Eq. \eqref{fermiGR2}, is still restricted by the Pauli principle: for $\omega_L \!\ll\! J/\hbar$ and when $E_{\text{F}}\!=\!-1.5 J$, this allows to probe the edge states that are located in the first bulk gap only. This important fact leads to the asymmetry highlighted in Figs. 4(c)-(d), which demonstrates the specific chirality of these edge states (cf. main text). 

We finally stress that the condition $$\hat h _{(-)}\!=\! \hat h _{(+)}\!=\!\hat{h}_0,$$ is necessary in order to probe the edge-state structure. Indeed, if we consider a simpler scheme in which the $(-)$ sector is no longer subjected to a synthetic gauge potential, we find that $N(q,\omega_L)-N(-q,\omega_L) \approx 0$. This observation shows that our scheme requires that the edge states of the $(-)$ sector should have the same chirality than the initially populated edge-states of the $(+)$ sector, \emph{i.e.} both systems should be subjected to the same magnetic flux (cf. Appendix D). \\

\section{Detection scheme using state-changing transitions: example for $^{171}$Yb atoms}

We give here a more detailed account of the detection scheme using $^{171}$Yb atoms (see Fig. 4 (a)). The ground $g$ and metastable excited $e$ states have zero electronic angular momentum but nuclear spin $I=1/2$. We denote the Zeeman manifolds $\{g_\downarrow,g_\uparrow \}$ and $\{e_\downarrow,e_\uparrow \}$ in the ground $^1 S_0$ and  $^3 P_0$ excited states, respectively. The states $g_\uparrow$ and $e_\uparrow$ are initially populated, as laser coupling between these two states is used to generate the artificial gauge field \cite{jaksch2003a,gerbier2010a} leading to Eq.~(1).  A crucial point is to ensure that topological edge states have the same structure in the initial and final states (cf. Appendix C). To this end, the initially unpopulated  states $(-)=\{g_\downarrow,e_\downarrow\}$ are also coupled by a laser generating the same gauge field as for $(+)$. The degeneracies are split by a relatively strong magnetic field, $\Delta E _{a_i} = - g_a m_i B$, where $a=e,g$ denotes the ground or excited manifold, $i=\uparrow/\downarrow$, $m_i=\pm 1/2$ the nuclear spin quantum number, $g_g/h\approx -750~$Hz/G, and $g_e/h\approx-1250$~Hz/G. A bias field $B\sim 100 ~$G thus leads to Zeeman shifts $\sim \pm 25~$kHz on the $\pi$ transitions and $\sim \pm 100~$kHz on the $\sigma^{\pm}$ transitions: these shifts are very large compared to the Rabi frequencies of both the gauge-field and probe lasers, which can thus be treated independently.

In order to probe the $(+)=\{g_\uparrow,e_\uparrow\}$ system, one introduces a weaker additional laser, coupling $g_\uparrow \rightarrow e_\downarrow$. Due to the gauge coupling, a population will build up in the $g_\downarrow$ state as well (roughly equal to that in the $e_\downarrow$ state since $\Omega \ll J$). Those atoms will be missing in the final detection step. After probing, the lattice sites are isolated by rapidly raising the lattice height and switching off the artificial gauge field. Atoms in the $g$ manifold are dispatched (possibly detected) using an auxiliary imaging transition $g \rightarrow f$. A natural choice for $^{171}$Yb is $f \!=$$^{1}\!P_1$, with a line width $\gamma_f/2\pi\approx 28~$MHz much larger than any Zeeman splitting in $g$ or $e$ (thus prohibiting independent detection of atoms depending on their spin $\downarrow /  \uparrow$). Crucially, atoms in the $e$ manifold are not in resonance with the imaging light and are therefore unaffected. The $e_\downarrow$ atoms are subsequently brought down to the $g$ manifold using, {\it e.g.}, adiabatic passage techniques, leaving the $e_{\uparrow}$ state unaffected. A further imaging pulse allows to detect those atoms, initially excited by the probe pulse. One might worry that a fraction of atoms from the $e_\uparrow$ state could end up being transferred too, thus contaminating the final edge signal. Fortunately, the off-resonant excitation rate to ``wrong" states will be smaller than the resonant rate by a factor scaling as $\sim (\Omega/\Delta_Z)^2$, with $\Delta_Z\sim \vert g_e-g_g \vert B/2$ a typical Zeeman splitting. Taking for example the parameters given in \cite{gerbier2010a}, one has $J/h\approx 100~$Hz, and $\Omega \sim 0.05 J/\hbar \sim 2\pi \times 20~$Hz, making the final contamination of $g_\downarrow$ by $e_\uparrow$ negligible ($\sim 10^{-5}$).

\begin{widetext}

\begin{table}[!]
\begin{tabular}{l | c | c  | l}
\hline 
\hline
Chern numbers associated to the two lowest bulk bands: & $N_{1}=-1$ & $N_{2}=2$   \\
Hall conductivity in the two bulk gaps  (from Kubo formula): & $\sigma_H (E_{\text{F}} \in 1\text{st gap})=-1$ & $\sigma_H (E_{\text{F}} \in 2\text{nd gap})=+1$   \\
Chirality of the edge states  inside the bulk gaps (circular geometry): & $\text{sign} (\sigma_H (1\text{st gap}))$=(-) & $\text{sign} (\sigma_H (2\text{nd gap}))$=(+)    \\
\hline
\hline
\end{tabular}
\caption{\label{cherntab} The Hofstadter model with $\Phi=1/3$: Chern numbers, Hall conductivity and edge-state configurations.}

\vspace{1cm}

\begin{tabular}{l | c | c  | c |  l}
\hline
\hline
Two-dimensional Geometries: & torus (abstract) & cylinder (abstract) & circular (realistic)  \\
\hline 
\hline
Number of 1D boundaries: & 0 & 2 & 1  \\
Chern number of the  bulk band $E_{\nu}(\bs{k})$: & $N_{\nu}$ & -- & --  \\
Hall conductivity for $E_{\text{F}} \in r$th bulk gap (from Kubo formula): & $\sigma_H^{(r)}=\sum_{\nu=1}^{r} N _{\nu}$ & -- & --  \\
\hline
Number of edge-state branches in the $r$th bulk gap: & -- & $2 \, \vert \sigma_H^{(r)} \vert$ & $\vert \sigma_H^{(r)} \vert$  \\
Chirality of the edge states  located in the $r$th bulk gap: & -- & $ \text{sign} (\sigma_H^{(r)} )$ & $\text{sign} (\sigma_H^{(r)} )$  \\
\hline
\hline
\end{tabular}
\caption{\label{bulkedgetab} The bulk-edge correspondence: Topological Chern numbers (defined in the bulk) and the properties of gapless edge excitations for the three geometries described in the text. Note that for the cylinder, the edge states at the two different physical edges have opposite velocities \cite{Hatsugai1993,Qi2006}.}
\end{table}
\end{widetext}

\end{document}